\def\fr#1#2{\hbox{${#1\over #2}$}}

\def\ni{\noindent}
\def\vs{\vskip.3cm}

\def\+{{(+)}}  \def\-{ {(-)} }   \def\0{ {(0)} }
\def\1{ {(1)} }  \def\2{ {(2)} }
\def\pd{\partial}

\def\con{\omega}

\def\sq{Q\kern-6pt/}
\def\sQ{Q\kern-12pt\nearrow}
\documentclass[11pt,eqs]{article}

\usepackage{latexsym,graphicx}

\def\be{\begin{equation}}             \def\ee{\end{equation}}
\def\ba{\begin{array}{rcl}}           \def\ea{\end{array}}
\def\beqa{\begin{eqnarray} }          \def\eeqa{\end{eqnarray} }
\def\beqalign{\begin{eqalign}}        \def\eeqalign{\end{eqalign}}
\def\leq#1{\label{eq:#1}}             \def\eq#1{(\ref{eq:#1})}
\def\bsubeq{\begin{subequations}}     \def\esubeq{\end{subequations}}
\def\bitem{\begin{itemize}}           \def\eitem{\end{itemize}}

\def\DJ{\leavevmode\setbox0=\hbox{D}\kern0pt
 \rlap{\kern.04em\raise.188\ht0\hbox{-}}D}
\def\dj{\leavevmode\setbox0=\hbox{d}\kern0pt
 \rlap{\kern.215em\raise.46\ht0\hbox{-}}d}

\newcommand{\bd}{\begin{displaymath}}
\newcommand{\ed}{\end{displaymath}}
\begin{document}

\title{Bosonic string theory in background fields by canonical methods
\thanks{Work supported in part by the Serbian Ministry of Science, Technology
and Development under contract No. 1486.}}
\author{B. Sazdovi\'c\thanks{e-mail address: sazdovic@phy.bg.ac.yu}\\
       {\it Institute of Physics, 11001 Belgrade, P.O.Box 57, Serbia}}
\maketitle
\begin{abstract}
We investigate classical dynamics of the bosonic string in the background
metric, antisymmetric and dilaton fields. We use canonical methods to find
Hamiltonian in terms of energy-momentum tensor components. The later are
secondary constraints of the theory. Due to the presence of the dilaton field
the Virasoro generators have nonlinear realization. We find that, in the curve
space-time, opposite chirality currents do not commute. As a consequence of
the two-dimensional general covariance, the energy-momentum tensor components
satisfy two Virasoro algebras, even in the curve space-time. We emphasize that
background antisymmetric and dilaton fields are the origin of space-time
torsion and space-time nonmetricity, respectively.
\end{abstract}
\vs

\ni {\it PACS number(s)\/}: 11.25.-w \par

\section{Introduction}

The string propagation in the flat Minkowski space-time has been investigated
for a long time and it is well understood \cite{GSW}. The dynamics of string
moving in arbitrary background metric, which can be interpreted as the
nonlinear $\sigma$-model, has also been studied \cite{FT,CFMP, BNS}. The goal
of these investigations was to analyze the consistency of string dynamics in
the presence of background fields. The condition for two-dimensional conformal
invariance on the quantum level, which technically means vanishing of the
renormalization group $\beta$-function for nonlinear $\sigma$-model, produce
the field equations for the background fields. The quantization of the theory
has been performed expanding the quantum fields $x^\mu$ around the expectation
value $x_0^\mu$, and using background field expansion by standard technique of
Riemann normal coordinates. This remarkable connection between the quantum
feature of the world-sheet (conformal invariance), and the classical
space-time property of the string theory (equations of motion), has been
obtained perturbatively.

In this paper, we are going to investigate string propagation in the arbitrary
gravitational, antisymmetric tensor, and dilaton background fields. We will
make classical {\it nonperturbative}  investigation of the {\it full
nonlinear} theory.

In Sec.2, we perform the canonical analysis, treating world-sheet fields as
variables in the theory, and space-time fields as a background depending on
the coordinates $x^\mu$. Because of the presence of the dilaton field, the
conformal component of the world-sheet metric survives, and corresponding new
currents appear. The energy-momentum tensor obtains the nonlinear realization.
It is not in the Sugawara form, but is a more general bilinear combination of
the currents, plus linear term in $\sigma$ derivative of the currents.

In Sec.3, Poisson bracket algebra of the currents and energy-momentum tensors
components has been investigated. Let us stress that, in spite of the fact
that some opposite chirality currents do not commute the energy-momentum
tensors components, which appear as secondary constraints,  satisfy two
Virasoro algebras as a consequence of the world-sheet reparametrization invariance.

We also derive the Hamiltonian equations of motion, in Sec.4, using the PB
relations between the energy-momentum tensor and the currents. With the help
of Lagrangian expressions for the currents we easy convert the above
Hamiltonian equations to the Lagrangian ones. Using these world-sheet
equations of motion, we introduce the space-time connection and covariant
derivative in the presence of the background fields.

In the conclusion, we discus the contribution of the antisymmetric tensor and
the dilaton field to the space-time torsion and nonmetricity, respectively.

The Appendix A is devoted to the world-sheet geometry.

\section{Canonical analysis of the theory}
\setcounter{equation}{0}

Let us consider bosonic string propagating in the non-trivial background,
described by the action \cite{GSW}-\cite{BNS}
\be
S= \kappa  \int_\Sigma d^2 \xi \sqrt{-g} \left\{ \left[ {1 \over 2}g^{\alpha\beta}G_{\mu\nu}(x)
+{\varepsilon^{\alpha\beta} \over \sqrt{-g}}  B_{\mu\nu}(x)\right]
\partial_\alpha x^\mu \partial_\beta x^\nu + \Phi(x) R^{(2)} \right\}  \, .   \leq{ac}
\ee
Here $\xi^\alpha \, (\alpha=0,1)$ are coordinates of two dimensional
world-sheet $\Sigma$ and $x^\mu(\xi) \,  (\mu =0,1,...,D-1)$ are coordinates
of the $D$ dimensional space-time $M$. The background space-time fields:
metric $G_{\mu \nu}$, antisymmetric tensor field $B_{\mu\nu}=- B_{\nu\mu}$ and
dilaton field $\Phi$, depend on the coordinates $x^\mu$. The intrinsic
world-sheet metric we denote by $g_{\alpha \beta}$ and corresponding scalar
curvature by $R^{(2)}$. Through the paper we will use the notation
$\partial_\alpha \equiv {\partial \over \partial \xi^\alpha}$ and
$\partial_\mu \equiv {\partial \over
\partial x^\mu}$.

If we parameterize the world-sheet metric tensor $g_{\alpha \beta }$ with the light-cone
variables  $(h^+,h^-, F)$ (for details see \eq{g} and  papers \cite{BPS,SM})
and introduce the useful combination of background fields
\be
\Pi_{\pm \mu \nu}= B_{\mu \nu} \pm {1 \over 2} G_{\mu \nu} \,   ,
\ee
we can write the action in the form
\be
S =  \kappa \int_\Sigma d^2 \xi \sqrt{-{\hat g}}\left[ 2 \Pi_{+ \mu \nu}{\hat \partial}_+ x^\mu
{\hat \partial}_- x^\nu + \Phi ({\hat R}^{(2)} + 2 {\hat \Delta} F ) \right] \,   ,  \leq{acP}
\ee
or more explicitly
\begin{eqnarray}
S&=&\kappa \int_\Sigma d^2 \xi {1 \over h^- -h^+}
\big[ G_{\mu \nu} ({\dot x}^\mu +h^- x^{\mu \prime }) ({\dot x}^\nu +h^+ x^{ \nu \prime })
-2 (h^- -h^+) B_{\mu \nu} {\dot x}^\mu x^{ \nu \prime }  \nonumber  \\
&+&2({\dot \Phi} +h^- {\Phi^\prime })({\dot F} +h^+ {F^\prime } +{h^{+ \prime }})
+2({\dot \Phi} +h^+ {\Phi^\prime })({\dot F} +h^- {F^\prime } +{h^{- \prime} }) \big] \, . \leq{ach}
\end{eqnarray}
Here in agreement with  \eq{vec} we introduce light-cone
derivatives
${\hat \partial}_\pm ={\sqrt{2} \over  h^- - h^+ }(\partial_0 +h^\mp \partial_1)$.
We also use the following notation ${\dot X} =\partial_\tau X$ and $ X^\prime =\partial_\sigma X$ for any variable $X$.

The momenta corresponding to the fields $x^\mu, h^\pm$ and $F$ are
\begin{eqnarray}
\pi_\mu &= {\partial {\cal L} \over \partial {\dot x}^\mu} = &{\kappa \over h^- -h^+} \big\{ G_{\mu \nu}
\big[2{\dot x}^\nu +(h^- +h^+) x^{\nu \prime}\big]
-2 (h^- -h^+) B_{\mu \nu} x^{\nu \prime }  \nonumber   \\
&&+2 \big[ 2{\dot F} +(h^- +h^+) {F^\prime } +{(h^-+h^+)^\prime} \big] \partial_\mu \Phi \big\}  \,  , \label{a}  \\
\pi_\pm &= {\partial {\cal L} \over \partial {\dot h}^\pm} = &0 \, ,  \label{b}  \\
\pi_F &= {\partial {\cal L} \over \partial {\dot F}} = &{2\kappa \over h^- -h^+}
\big[ 2{\dot \Phi} +(h^- +h^+) {\Phi^\prime } \big] \,  .   \label{c}   \leq{2.3}
\end{eqnarray}
From the eq. (\ref{c}) we can solve the time derivatives of the field $\Phi$ as
\be
{\dot \Phi}= {1 \over 4 \kappa} (h^- i_-^\Phi - h^+ i_+^\Phi)  \,  ,  \leq{fit}
\ee
where we introduced the currents
\be
i_\pm^\Phi= \pi_F \pm 2\kappa \Phi' \,  .   \leq{2.5}
\ee

We will denote the gradient of the dilaton field $\Phi$ by $a_\mu$
\be
a_\mu = \partial_\mu \Phi \, , \qquad a^\mu =G^{\mu \nu} a_\mu   \, , \qquad
a^2=G^{\mu \nu} a_\mu a_\nu  \,  .          \leq{ami}
\ee
We will suppose that the gradient of the dilaton field $\Phi(x)$ is not
light-cone vector, and the condition $a^2 \neq 0$ is fulfilled in the whole
space-time.

Multiplying eq. (\ref{a}) with $a^\mu$ we can solve it in terms of ${\dot F}$
\be
{\dot F}= {1 \over 4 \kappa} (h^- i_-^F - h^+ i_+^F)-{1 \over 2}
{ (h^-+h^+)^\prime}    \,  ,       \leq{Ft}
\ee
where we define new currents
\be
i_\pm^F= {a^\mu \over a^2} j_{\pm \mu}-{1 \over 2 a^2} i_\pm^\Phi
\pm 2 \kappa {F^\prime }  \,  ,   \leq{2.8}
\ee
and
\be
j_{\pm \mu} =\pi_\mu +2\kappa \Pi_{\pm \mu \nu} {x^\nu}' \,  .      \leq{jmi}
\ee
We used the relation ${\dot \Phi} = a_\mu {\dot x}^\mu$, which allow us to
trade ${\dot x}^\mu$ for ${\dot \Phi}$. Substituting the
expression for ${\dot F}$ back into (\ref{a}), we can solve the same equation
in terms of  ${\dot x}^\mu$
\be
{\dot x}^\mu =  {G^{\mu \nu} \over 2 \kappa} (h^- J_{- \nu} - h^+ J_{+ \nu}) \,  .  \leq{xt}
\ee
The corresponding currents $J_{\pm \mu}$ are defined as
\be J_{\pm \mu} =P^T{}_\mu{}^\nu j_{\pm \nu} +{a_\mu \over 2
a^2}i_\pm^\Phi = j_{\pm \mu} -{ a_\mu \over a^2} j \,  ,  \leq{J}
\ee
and we introduce the variable
\be
j=a^\mu j_{\pm \mu} -{1 \over 2} i_\pm^\Phi =a^2 (i_\pm^F \mp 2 \kappa F^\prime)  \,   .   \leq{j}
\ee

The projection operator
\be
P^T{}_{\mu \nu} = G_{\mu \nu} - {a_\mu a_\nu \over a^2} \,  ,   \leq{PT}
\ee
will arise frequently in this article. Let us explain its geometrical
interpretation. Note that $a_\mu$ is a normal vector to the $D-1$ dimensional
submanifold $M_{D-1}$, defined by the condition $\Phi(x)= const$. For $a^2
\neq 0$, the corresponding unit vector  is $n_\mu = {a_\mu \over
\sqrt{\varepsilon a^2}}$. Here $\varepsilon=1$ if $a_\mu$ is time like vector
($a^2 > 0$), and $\varepsilon=-1$ if $a_\mu$ is space like vector ($a^2 < 0$),
so that $n^2 =\varepsilon$. Consequently, we can interpret the expression
\be
P^T{}_{\mu \nu} = G_{\mu \nu}- \varepsilon n_\mu n_\nu \equiv G_{\mu \nu}^{D-1} \,  ,  \leq{im}
\ee
as a {\it induced metric} on $M_{D-1}$.

The time derivatives of the fields $\Phi , F$ and $x^\mu$ we expressed in terms of the corresponding
currents \eq{fit},\eq{Ft} and \eq{xt}. The $\sigma$ derivative of the same fields we can express
in terms of the same currents
\be
x^{\mu \prime}= {G^{\mu \nu} \over 2 \kappa} (J_{+ \nu} - J_{- \nu})  \, , \qquad
{F^\prime }= {1 \over 4 \kappa} (i_+^F - i_-^F)   \,  ,  \qquad
{\Phi^\prime }= {1 \over 4 \kappa} (i_+^\Phi - i_-^\Phi)  \,   .   \leq{prim}
\ee

Note that we introduce the momenta $\pi_\mu , \pi_\pm$ and $\pi_F$ for two
dimensional fields $x^\mu, h^\pm$ and $F$, respectively. We did not introduce
momentum for dilaton background field $\Phi$, but for convenience we have the
currents $i_\pm^\Phi$ in term of which we expressed the ${\dot \Phi}$ and
$\Phi^\prime$. These variables are not independent, because of the relations
${\dot \Phi}= a_\mu {\dot x}^\mu$ and ${\Phi^\prime}= a_\mu x^{\mu \prime}$,
which are equivalent to the following connection between the currents
$i_\pm^\Phi =2 a^\mu J_{\pm \mu}$. The last one can be confirmed directly from
\eq{J}.

After long but straightforward calculation, the canonical Hamiltonian density
obtains the form
\be
{\cal H}_c= h^- T_- + h^+ T_+ + {q^\prime }  \,   ,  \leq{hc}
\ee
where the expressions for the energy momentum tensor components
\be
T_\pm =\mp {1 \over 4\kappa} \left(G^{\mu\nu} J_{\pm \mu} J_{\pm \nu} + i_\pm^F i_\pm^\Phi \right)
+{1 \over 2} i_\pm^{\Phi \prime} =
\mp {1 \over 4\kappa} \left( G^{\mu\nu}
j_{\pm \mu} j_{\pm \nu} -{j^2 \over a^2} \right) +i_\pm  \,   ,  \leq{emt}
\ee
contains the useful combination of the currents
\be
i_\pm= {1 \over 2} ({i_\pm^\Phi}' - F' i_\pm^{\Phi}) \,  .  \leq{2.16}
\ee
In the expression for the canonical Hamiltonian $H_c=\int d \sigma {\cal H}_c $ the term
\be
q= -{1 \over 2} (h^- i_-^\Phi + h^+ i_+^\Phi) \,   ,    \leq{2.17}
\ee
appears only on the world-sheet boundary $\partial \Sigma$ and does not contribute to the
equations of motion.

\section{Virasoro algebra in the curved space-time}
\setcounter{equation}{0}

We shall start with basic Poisson brackets (PB) algebra
\be
\{ x^\mu , \pi_\nu \} = \delta^\mu_\nu \delta  \,  .  \leq{PB0}
\ee
Through the paper, we will understand that the first variable in PB depends on
$\sigma$ and the second one on ${\bar \sigma}$. We will also use the
conventions $\delta \equiv \delta(\sigma - {\bar \sigma})$, $\delta^{\prime}
\equiv \partial_\sigma \delta(\sigma - {\bar \sigma})$ and $\delta^{\prime
\prime} \equiv \partial_\sigma^2 \delta(\sigma - {\bar \sigma})$. On the right
hand side, all ${\bar \sigma}$ dependence we will  write explicitly and
$\sigma$ dependence we will omit.

Let us first calculate PB between the currents $j_{\pm \mu}$, $i_\pm^\Phi$, $j$ and $i_\pm$
\be
\{j_{\pm \mu} , j_{\pm \nu} \}= \pm 2 \kappa (G_{\mu \nu} { \delta^\prime}
+ \Gamma_{\mp \mu , \nu \rho} x^{\rho \prime } \delta) \,  , \qquad
\{j_{\pm \mu} , j_{\mp \nu} \}= \pm 2 \kappa \Gamma_{\mp \rho , \mu \nu} {x^\prime }^\rho \delta  \,  ,    \leq{PB1}
\ee
\be
\{j_{\pm \mu} , i_\pm^\Phi \}= \pm 2 \kappa a_\mu { \delta^\prime} \,  , \qquad
\{j_{\pm \mu} , i_\mp^\Phi \}= \mp 2 \kappa a_\mu {\delta^\prime} \, ,  \leq{PB2}
\ee
\be
\{i_\pm^\Phi , i_\pm^\Phi \} =0 \,  , \qquad
\{ i_\pm^\Phi ,i_\mp^\Phi \} =0 \,   ,     \leq{PB3}
\ee
\be
\{ j ,j \} =0  \,  ,  \qquad
\{ j , i_\pm^\Phi \} = \pm 2 \kappa a^2 { \delta^\prime} \, , \qquad
\{ j , i_\pm \}=\mp \kappa a^2 [\delta^{\prime \prime}+F^\prime({\bar \sigma})\delta^{\prime }]
+{1 \over 4}i_\pm^\Phi ({\bar \sigma})\delta^{\prime } \,  ,   \leq{PB4}
\ee
\be
\{ i_\pm , i_\pm \} =-  {1 \over 2} [i_\pm(\sigma) + i_\pm({\bar \sigma}) ]
{ \delta^\prime} \, ,  \qquad
\{ i_\pm , i_\mp \} = - {1 \over 2} [i_\mp(\sigma) + i_\mp({\bar \sigma}) ] {\delta^\prime}
\pm \kappa {\Phi^\prime} ({\delta^{\prime \prime}} +{F^\prime } { \delta^\prime} +
{ F^{ \prime \prime}} \delta )  \, ,      \leq{PB5}
\ee
\be
\{ i_\pm , i_\pm^\Phi \} =- {1 \over 2} i_\pm^\Phi { \delta^\prime} \,  , \qquad
\{ i_\pm , i_\mp^\Phi \} =- {1 \over 2} i_\pm^\Phi { \delta^\prime} \,  .   \leq{PB6}
\ee
The expressions $\Gamma_{\pm \rho , \mu \nu}$  are defined as
\be
 \Gamma^\rho_{\pm \nu \mu}= \Gamma^\rho_{\nu \mu} \pm B^\rho_{\nu \mu}    \,  ,  \leq{coB}
\ee
where the first term is Christoffel connection
\be
\Gamma^\rho_{\nu \mu}={1 \over 2} G^{\rho \sigma}(\partial_\nu G_{\mu \sigma}
+ \partial_\mu G_{\nu \sigma}- \partial_\sigma G_{\nu \mu})   \,  ,        \leq{coG}
\ee
and the second one is the antisymmetric tensor field strength
\be
B_{\mu \nu \rho}= \partial_\mu B_{\nu \rho} + \partial_\nu B_{\rho \mu} + \partial_\rho B_{\mu \nu}=
D_\mu B_{\nu \rho} + D_\nu B_{\rho \mu} + D_\rho B_{\mu \nu}   \,  .       \leq{fsB}
\ee
The "generalized connection" is expressed in terms of the variables $G_{\pm \mu \nu} = \pm 2
\Pi_{\pm \mu \nu}= G_{\mu \nu} \pm 2 B_{\mu \nu}$ in the similar way as Christoffel connection is
expressed in terms of $G_{\mu \nu}$
\be
\Gamma_{\pm \rho, \nu \mu}= \frac{1}{2} (\partial_\nu G_{\pm \mu \rho}
+\partial_\mu G_{\pm \rho \nu}-\partial_\rho G_{\pm \mu \nu}) \,  .       \leq{coP}
\ee

For any momenta independent variables, $X$, we have
\be
\{ X , T_\pm \} =\mp {1 \over 2 \kappa} J^\mu_\pm \partial_\mu X \delta  \,  . \leq{PB7}
\ee

Let us write the expression for the energy-momentum tensor components in the
form $T_\pm= t_\pm \pm t$, where
\be
t_\pm= \mp {1 \over 4 \kappa} G^{\mu \nu} j_{\pm \mu} j_{\pm
\nu} + i_\pm \, , \qquad  t= {1 \over 4 \kappa} { j^2 \over a^2}  \,  ,  \leq{t}
\ee
and let us introduce the combination
\be
\tau= {1 \over 2}( { j^\prime} -{ F^\prime} j ) \,  .  \leq{tau}
\ee
Then we obtain the following algebra
\be
\{ t_\pm , t_\pm \}= -[ t_\pm(\sigma) +t_\pm({\bar \sigma}) ] { \delta^\prime}
- [ \tau(\sigma) +\tau({\bar \sigma}) ] {\delta^\prime} \,  ,  \leq{vir}
\ee
\be
\{ t_\pm , t \} +\{ t , t_\pm \}= -[ t(\sigma) +t({\bar \sigma}) ] { \delta^\prime}
\pm [ \tau(\sigma) +\tau({\bar \sigma}) ] { \delta^\prime} \, ,
\ee \be \{t ,t \} =0 \,  ,
\ee
which produce the Virasoro algebra for the same chirality energy-momentum tensor components
\be
\{ T_\pm , T_\pm \}= -[ T_\pm(\sigma) +T_\pm({\bar \sigma}) ] { \delta^\prime} \,   .   \leq{Vir}
\ee
Similarly, for the opposite chiralities we have
\be
\{ t_\pm , t_\mp \}= [ \tau(\sigma) +\tau({\bar \sigma}) ] {\delta^\prime} \,  , \leq{virpm}
\ee
\be
\{ t_\pm , t \} -\{ t , t_\pm \}= \pm [ \tau(\sigma) +\tau({\bar \sigma}) ]
{ \delta^\prime} \, ,
\ee
which contribute to the fact that the opposite chirality energy-momentum tensor
components commute
\be
\{ T_\pm , T_\mp \}= 0    \,  .    \leq{Virpm}
\ee
Consequently, the expression \eq{emt} presents new, {\it non-linear}
realization of the Virasoro algebras. The dilaton
field $\Phi$ is the origin of the non-linearity. We want to stress, that the
opposite chirality currents generally do not commute in the curved space time,
see \eq{PB1}-\eq{PB6}, but the opposite chirality energy-momentum tensors
components commute.

The momenta $\pi_\pm$, defined in (\ref{b}), are primary constraints. The total
Hamiltonian takes the form $H_T = \int d \sigma ( {\cal H}_c +\lambda^- \pi_-
+\lambda^+ \pi_+)$. The consistency conditions ${\dot \pi}_\pm = \{ \pi_\pm ,
H_T \} =-T_\pm $ produce the secondary constraints $T_\pm$. As a consequence
of the Virasoro algebra \eq{Vir} and \eq{Virpm}, the consistency conditions
for the secondary constraints $T_\pm$ are satisfied and there are no more
constraints. All constraints are first class and they are generators of two
dimensional diffeomorphisms \cite{BPS}. The canonical Hamiltonian is weakly
equal to zero.

\section{Equations of motion}
\setcounter{equation}{0}

Using the expressions for the energy momentum tensor and for the currents
obtained in the previous subsections, we are going to find equations of motion
for the action \eq{ac}. First we obtain Hamiltonian equations of motion and
then the corresponding Lagrangian ones. It is useful to define the expression
$V_\pm = V^\mu(x) J _{\pm \mu}$, where $V^\mu(x)$ is an arbitrary vector
function of the coordinates $x^\mu$. Then with the help of \eq{hc} and
\eq{emt} we have
\be
{\dot V}_\pm = \{ V_\pm , H_c \}= - (h^\pm V_\pm)^\prime + {h^- - h^+  \over 2 \kappa}
({}^\star D_{\mp \mu} V_{\nu}) J_\pm^\nu J_\mp^\mu   \, ,   \leq{dotV}
\ee
where we introduce notation
\begin{eqnarray}
&&{}^\star D_{\pm \mu} V_{\nu} =\partial_\mu V_{\nu} -
{}^\star \Gamma^\rho_{\pm \nu \mu}  V_{\rho} =
 D_{\pm \mu} V_{\nu} -{a^\rho  V_{\rho} \over  a^2} D_{\pm \mu} a_\nu \,  , \nonumber  \\
{}^\star \Gamma^\rho_{\pm \nu \mu}=&& \Gamma^\rho_{\pm \nu \mu} +{a^\rho \over a^2} D_{\pm \mu} a_\nu
= P^{T \rho}{}_\sigma \Gamma^\sigma_{\pm \nu \mu} +
{a^\rho \over a^2} \partial_\mu a_\nu = \Gamma^\rho_{\nu \mu} \pm P^{T \rho}{}_\sigma B^\sigma_{\nu \mu}
+{a^\rho \over a^2} D_{\mu} a_\nu   \,  .  \quad   \leq{cdc}
\end{eqnarray}

Under space-time general coordinate transformations the expression ${}^\star
\Gamma^\rho_{\pm \nu \mu}$ transforms as a connection. It contains the
Christoffel connection \eq{coG}, itself. If we consider ${}^\star
\Gamma^\rho_{\pm \nu \mu}$ as a space-time connection, then ${}^\star D_{\pm
\mu}$ is corresponding covariant derivative. The consequences of such approach
we will explain in detail in the next paper \cite{BS}.

We can rewrite \eq{dotV} in the world-sheet covariant way
\be
{\hat \nabla}_\mp V_\pm = {1 \over \sqrt{2} \kappa}
({}^\star D_{\mp \mu} V_{\nu}) J_\pm^\nu J_\mp^\mu   \, .  \leq{emV}
\ee
For $V_\mu = const$, the last equation produce the equation of motion for the current $J_\pm^\mu$
\be
{\hat \nabla}_\mp J_\pm^\mu + {1 \over \sqrt{2} \kappa}
{}^\star \Gamma_{\mp \rho \sigma}^\mu  J_\pm^\rho J_\mp^\sigma =0   \, ,   \leq{emJ}
\ee
and for $V_\mu = a_\mu$, the equation of motion for the current $i_\pm^\Phi$
\be
{\hat \nabla} _\mp i_\pm^\Phi =0  \,  .    \leq{emifi}
\ee
From \eq{emV} and \eq{emJ} we also obtain
\be
{\hat \partial}_\mp V_\mu = {1 \over \sqrt{2} \kappa} \partial_\nu V_\mu J_\mp^\nu \,  .  \leq{isf}
\ee
Similarly, the equation of motion for the current $i_\pm^F$
has the form
\be
{\hat \nabla}_\mp i_\pm^F  \pm {2 \sqrt{2} \kappa \over h^- -h^+} h^{\pm \prime \prime}
 = {1 \over \sqrt{2} \kappa a^2} (D_{\mp \mu} a_\nu) J_\pm^\nu J_\mp^\mu   \, . \leq{emiF}
\ee
So, \eq{emJ}, \eq{emifi} and \eq{emiF} are Hamiltonian equations of motion.
The equations for the momenta $\pi_\pm$ produce the vanishing of the energy
momentum tensor components $T_\pm =0$.

To obtain corresponding Lagrangian equations, it is possible to substitute the
expressions for the momenta into the Hamiltonian equations, but there is
equivalent and simpler approach. First we obtain the Lagrangian expressions
for the currents
\be
J_{\pm \mu}= \sqrt{2} \kappa G_{\mu \nu} {\hat \partial}_\pm x^\nu \,  ,  \qquad
i_{\pm}^\Phi= 2 \sqrt{2} \kappa  {\hat \partial}_\pm \Phi \, , \qquad
i_{\pm}^F=  \sqrt{2} \kappa ( 2 {\hat \partial}_\pm F \mp {\hat \con}_\pm \pm {\hat \con}_\mp ) \,  ,
\ee
solving linear system of the equations, \eq{xt}, \eq{fit}, \eq{Ft} and
\eq{prim} and eliminating $\tau$ and $\sigma$ derivatives of the corresponding
variables. Then, we substitute them into Hamiltonian equations and find
\be
{\hat \Delta} x^\mu -2 {}^\star \Gamma_{\mp \rho \sigma}^\mu  {\hat \partial}_\pm x^\rho
{\hat \partial}_\mp x^\sigma =0  \,  ,           \leq{lemJ}
\ee
\be
{\hat \Delta} \Phi =0  \,      ,    \leq{lemfi}
\ee
\be
{\hat \Delta} F + {1 \over 2} {\hat R}^{(2)} + {1 \over a^2} (D_{\mp \mu} a_\nu)
{\hat \partial}_\pm x^\nu {\hat \partial}_\mp x^\mu =0  \,  ,     \leq{lemF}
\ee
\be
T_\pm = \mp \frac{\kappa}{2} (G_{\mu \nu}  {\hat \partial}_\pm x^\mu {\hat \partial}_\pm x^\nu
-2 {\hat \nabla}_\pm {\hat \partial}_\pm \Phi +4 {\hat \partial}_\pm F {\hat \partial}_\pm \Phi - {\hat \Delta} \Phi )=0  \,  .  \leq{lemh}
\ee

Note that equations \eq{lemJ} (with $+$ and $-$ indices) are equivalent, because of the property
${}^\star \Gamma_{\pm \rho \sigma}^\mu = {}^\star \Gamma_{\mp \sigma \rho}^\mu$, as well as
equations \eq{lemF} are equivalent, because of the property $D_{\pm \mu} a_\nu = D_{\mp \nu} a_\mu$.

The Lagrangian expression for the equation \eq{isf} produce just the chain rule
\be
{\hat \partial}_\pm V_\mu = {\hat \partial}_\pm x^\nu \partial_\nu V_\mu \,     .
\ee
As a consequence of the relation $i_\pm^\Phi = 2 a^\mu J_{\pm \mu}$, the
equation \eq{lemfi} follows from \eq{lemJ} and can be obtained just
multiplying the last one by $a_\mu$.

Finally, we can go from the hat variables to the original ones (see App. A)
and obtain the final form of the Lagrangian field equations
\be
\nabla_\mp \partial_\pm  x^\mu + {}^\star \Gamma_{\mp \rho \sigma}^\mu
\partial_\pm x^\rho \partial_\mp x^\sigma =0  \,  ,    \leq{lJ}
\ee
\be
G_{\mu \nu}  \partial_\pm  x^\mu \partial_\pm x^\nu -2 \nabla_\pm
\partial_\pm \Phi =0   \,  ,  \leq{lh}
\ee
\be
R^{(2)} + {2 \over a^2} (D_{\mp \mu} a_\nu) \partial_\pm  x^\nu \partial_\mp  x^\mu =0
  \,  .     \leq{lF}
\ee

\section{Conclusions and discussion}
\setcounter{equation}{0}

In this paper, we investigated bosonic string theory in presence of
background fields. We used the Hamiltonian method and treated the space-time
coordinates $x^\mu$ and the world-sheet metric $g_{\alpha \beta}$ as variables
in the theory. We did the analysis for the arbitrary background fields as
functions of the string coordinates.

We obtained a new, nonlinear realization of the energy-momentum tensor
components \eq{emt}. They satisfy two independent Virasoro algebras, which is
expected because the theory is invariant under world-sheet reparametrization.
These components are bilinear in the currents or linear in the $\sigma$
derivative of the currents. All currents are linear in the momenta and in the
$\sigma$ derivative of the fields. Let us stress that the opposite chirality
components of the energy-momentum tensor commute, \eq{Virpm}, in spite of the
fact that the corresponding currents do not commute. Using appropriate
commutation relations we obtain the Hamiltonian equations of motion which
easily produce the Lagrangian ones.

The equations of motion \eq{lJ}-\eq{lF} offer the possibility to investigate
the dynamics of the string, propagating in the curved target space. Starting
with the given stringy connection  \eq{cdc}, let us here shortly discuss the
main features of the target space geometry. The space-time observed by the
string moving in the background $G_{\mu \nu}$, $B_{\mu \nu}$ and $\Phi$, we
will call stringy space-time.

The antisymmetric part of the stringy connection is the {\bf stringy torsion}
\be
{}^\star T_\pm{}^\rho_{\mu \nu} \equiv {}^\star \Gamma_\pm{}^\rho_{\mu \nu} - {}^\star \Gamma_\pm{}^\rho_{\nu \mu} =
\pm 2 P^{T \rho}{}_\sigma  B^\sigma_{\mu \nu}         \,  ,   \leq{T}
\ee
which is the transverse part of the field strength of the antisymmetric tensor
field $B_{\mu \nu}$.

We define the stringy parallel transport as usual
\be
{}^\star \delta_\pm V^\mu = - {}^\star \Gamma_{\pm \rho \sigma}^\mu V^\rho d x^\sigma  \,  .  \leq{ptcn}
\ee
The metric postulate is defined by the demand, that after parallel transport
the metric is equal to the local one. The most interesting feature of the
stringy geometry is the breaking of the space-time metric postulate. The {\bf
stringy nonmetricity}
\be
{}^\star Q_{\pm \mu \rho \sigma} \equiv  -{}^\star D_{\pm \mu} G_{\rho \sigma} =
{1 \over a^2} D_{\pm \mu} (a_\rho a_\sigma )       \,   ,   \leq{mp}
\ee
is different from zero, which means that metric $G_{\mu \nu}$ is not compatible with stringy connection
${}^\star \Gamma^\mu_{\pm \nu \rho}$.
Consequently, during parallel transport the length and angle deformations depend on the vector field $a_\mu$.

So, with the help of the equations of motion, we recognize the expressions
${}^\star \Gamma^\rho_{\pm \mu \nu}$ as stringy connections. It helps us to
conclude that the string can feel the space-time torsion and nonmetricity. The
more detailed investigation of the stringy geometry and geometrical
interpretation of the equations of motion will be presented in \cite{BS}.

\appendix 

\section{World-sheet geometry in the light-cone basis}
\setcounter{equation}{0}

In this appendix we present our notations and conventions concerning
intrinsic world sheet geometry in the light-cone basis.

We parameterize the world-sheet metric $g_{\alpha \beta }$ with light-cone
variables  $(h^+, h^-, F)$  \cite{BPS,SM}
\be
g_{\alpha \beta} =e^{2F} {\hat g}_{\alpha \beta}=
\fr{1}{2}e^{2F}\pmatrix{ -2h^-h^+    &  h^-+h^+ \cr
           h^-+h^+    &  -2      \cr }\, .               \leq{g}
\ee
The interval of the world-sheet $\Sigma$
\be
ds^2 = g_{\alpha \beta} d \xi^\alpha d \xi^\beta = 2 d \xi^+ d \xi^-  \,   ,
\ee
can be expressed in terms of the light-cone coordinates
\be
d \xi^\pm = { \pm 1 \over \sqrt{2}} e^F ( d \xi^1 - h^\pm d \xi^0) = e^F d {\hat \xi}^\pm  = e^\pm{}_\alpha  d \xi^\alpha \,    .
\ee
The quantities $ e^\pm{}_\alpha$ define the light-cone one form basis
$\theta^\pm = e^\pm{}_\alpha d \xi^\alpha$, and its inverse define the tangent
vector basis $e_\pm = e_\pm{}^\alpha \partial_\alpha = \partial_\pm$. In this
basis, components of the arbitrary vector $V_\alpha$ are
\be
V_{\pm}=e^{-F} {\hat V}_\pm =e_{\pm}{}^\alpha V_\alpha =
{\sqrt{2} e^{-F} \over h^- -h^+} (V_0+h^{\mp}V_1)\,  .     \leq{vec}
\ee
The world-sheet covariant derivatives on tensor $X_n$ are
\be
\nabla_\pm X_n = (\pd_{\pm} +n \con_{\pm}) X_n   \,  ,        \leq{4.6}
\ee
where the number $n$ is sum of the indices, counting index $+$ with $1$ and
index $-$ with $-1$, and
\be
\con_{\pm} =e^{-F}({\hat \con}_\pm \mp {\hat \partial}_\pm F) \,  , \qquad
{\hat \con}_\pm =\mp {\sqrt{2}\over h^- -h^+} h^{\mp \prime} \,  ,
\ee
are two dimensional Riemannian connections. In this notations the world-sheet
scalar curvature has the form
\be
R^{(2)} =2 \nabla_- \con_+ -2 \nabla_+ \con_-      \,   .          \leq{R2}
\ee
We also use the relation
\be
\sqrt{-g} R^{(2)} =\sqrt{-{\hat g}}({\hat R}^{(2)} +2 {\hat \Delta}F) \,   ,        \leq{RR}
\ee
where
\be
{\hat \Delta}=-2 {\hat \nabla}_\pm {\hat \partial}_\mp   \,  ,  \leq{L}
\ee
is the Laplace operator.

\end{document}